\documentstyle[prl,twocolumn,aps]{revtex}
\input{psfig}
\def\apj #1 #2 #3 {#1, ApJ, {\bf #2}, #3}
\def\apjl #1 #2 #3 {#1, ApJ, {\bf #2}, L#3}
\def\apjs #1 #2 #3 {#1, ApJS, {\bf #2}, #3}
\def\aap  #1 #2 #3 {#1, A\&A, {\bf #2}, #3}
\def\mnras #1 #2 #3 {#1, MNRAS, {\bf #2}, #3}
\def\pra #1 #2 #3 {#1, Phys.~Rev.~A., {\bf #2}, #3}
\def\prb #1 #2 #3 {#1, Phys.~Rev.~B., {\bf #2}, #3}
\def\prc #1 #2 #3 {#1, Phys.~Rev.~C., {\bf #2}, #3}
\def\prd #1 #2 #3 {#1, Phys.~Rev.~D., {\bf #2}, #3}
\def\pre #1 #2 #3 {#1, Phys.~Rev.~E., {\bf #2}, #3}
\def\prl #1 #2 #3 {#1, Phys.~Rev.~Lett., {\bf #2}, #3}
\def\plb #1 #2 #3 {#1, Phys.~Lett.~B., {\bf #2}, #3}
\def\science #1 #2 #3 {#1, Science., {\bf #2}, #3}
\def\nature #1 #2 #3 {#1, Nature., {\bf #2}, #3}
\def\nphysa #1 #2 #3 {#1, Nucl.~Phys.~A., {\bf #2}, #3}
\def\nphysb #1 #2 #3 {#1, Nucl.~Phys.~B., {\bf #2}, #3}
\def\nphysbs #1 #2 #3 {#1, Nucl.~Phys.~B.~Suppl., {\bf #2}, #3}

\def\h#1{\hbox{${}^{#1}$H}}

\def\omegab{\hbox{${\sl\Omega}_{\rm b}$}}
\def\h502{\hbox{$ h^{2}_{50}$}}
\def\xinue{\hbox{$\xi_{\nu_{e}}$}}
\def\xinum{\hbox{$\xi_{\nu_{\mu}}$}}
\def\xinut{\hbox{$\xi_{\nu_{\tau}}$}}
\def\xinumt{\hbox{$\xi_{\nu_{\mu , \tau}}$}}
\def\nue{\hbox{$\nu_{e}$}}
\def\num{\hbox{$\nu_{\mu}$}}
\def\nut{\hbox{$\nu_{\tau}$}}

\def\fun#1#2{\lower3.6pt\vbox{\baselineskip0pt\lineskip.9pt
  \ialign{$\mathsurround=0pt#1\hfil##\hfil$\crcr#2\crcr\sim\crcr}}}
\begin{document}
\bigskip
\bigskip
%

\title{Constraints on Neutrino Degeneracy from the Cosmic Microwave Background
and Primordial Nucleosynthesis}
\author{ M. Orito$^{1}$, T. Kajino$^{1}$, G. J. Mathews$^{2}$,
 Y. Wang$^{3}$
}
\address{$^1$National Astronomical Observatory, Mitaka, Tokyo 181-8588, Japan
}
\address{$^2$University of Noter Dame, Center for Astrophysics, Notre Dame, IN 46556
}
\address{$^3$University of Oklahoma, Department of Physics \&  Astronomy,
Norman, OK  73019
}
\date{\today}
\maketitle
\begin{abstract}
We reanalyze the cosmological constraints on the existence of
a net universal lepton asymmetry and neutrino degeneracy
based upon the latest high resolution CMB sky maps 
from BOOMERANG, DASI, and MAXIMA-1.  We generate 
likelihood functions by marginalizing over 
$(\omegab h^2,\xinumt, \xinue, \Omega_\Lambda,h,n)$ plus the calibaration 
uncertainties.  We consider flat $\Omega_M + \Omega_\Lambda = 1$
cosmological models with two identical degenerate neutrino
species, $\xinumt \equiv  \vert \xinum \vert = \vert \xinut \vert$
and a small $\xinue$.
We assign weak top-hat priors on the electron-neutrino degeneracy parameter
$\xinue$ and $\Omega_b h^2$ based upon allowed values
consistent with the nucleosynthesis constraints as a function
of $\xinumt$.  The change in the
background neutrino temperature with degeneracy is also explicitly included,
and Gaussian priors for $h = 0.72 \pm 0.08$  and the experimental calibration
uncertainties are adopted.
The marginalized likelihood functions 
show a slight (0.5$\sigma$) preference for neutrino degeneracy.
Optimum values with two equally degenerate $\mu$ and $\tau$ neutrinos
imply $\xinumt = 1.0^{+0.8 (1\sigma)}_{-1.0 (0.5\sigma)}$,
 from which we deduce $\xi_{\nu_e} =  0.09^{+0.15}_{-0.09}$, and $\Omega_b h^2
=  0.021^{+0.06}_{- 0.002}$.
The $2\sigma$ upper limit becomes 
$ \xinumt \le 2.1$, which implies 
$\xi_{\nu_e} \le 0.30$, and $\Omega_b h^2 \le 0.030$.                    
For only
a single large-degeneracy species  the optimal value is 
$\vert \xinum \vert$ or $\vert \xinut \vert = 1.4$ with
a $2\sigma$ upper limit of
 $\vert \xinum \vert$ or $\vert \xinut \vert \le 2.5$
\end{abstract}
\section{Introduction}
The present relic neutrino number asymmetry is not directly observable.  Hence,
there is no firm experimental basis for postulating that the lepton number
for each species is zero. Charge neutrality, however, demands that any
universal net lepton number beyond the
net baryon number must reside entirely within the neutrino sector.  
It has been suggested that 
the total lepton number could be large in the context
of the $SU(5)$ and $SO(10)$ grand unified theories 
\cite{harvey81,fry82,dolgov91,dolgov92},
or supersymmetric baryogenesis
\cite{casas99,mcd00,king00} based upon the Affleck-Dine scenario
\cite{affleck85}.  Such mechanisms  might 
generate lepton number asymmetry up to ten orders of magnitude 
larger than the baryon number
asymmetry. Furthermore, even if one demands that $B-L \approx 0$,
it is possible for the lepton
numbers $L_l$ of individual neutrino species to
be large compared to the baryon number of the
universe, $B$, as long as the net total lepton number is small.

Moreover, there presently exists at least some marginal
cosmological evidence for neutrino asymmetry.
For example, neutrinos with large lepton asymmetry and masses
$\sim 0.07$ eV might be required to explain the 
existence of cosmic rays  with energies in excess of
the Greisen-Zatsepin-Kuzmin cutoff \cite{greisen66,gelmini99}.
Also, degenerate, massive (2.4 eV) neutrinos might 
be required \cite{larsen95}
to provide a good fit to the power spectrum of large scale
structure in mixed dark matter models.
It is thus important to carefully scrutinize
the limits which cosmology places on the allowed 
values of possible universal lepton asymmetry.
Indeed, a number of recent papers 
\cite{hu,lineweaver,kinney,lesgourgues,hannestad,lesgourgues00,lesgourguesm,kneller,esposito,hansen01,bowen,dolgov02}
have addressed this issue with varying degrees of complexity.
The present work differs from those in several details as summarized
below.  It represents an independent examination of this issue.

\section{Present Approach}
In a recent paper \cite{orito00} we considered new constraints imposed
on neutrino degeneracy from primordial nucleosynthesis.
Particular attention was paid to the neutrino decoupling
temperatures before the nucleosynthesis epoch.
Of relevance to the present work is that we have shown that
neutrinos can decouple at a higher temperature than estimated in earlier
studies \cite{kang92}.  
This means that more particle
degrees of freedom could be present at neutrino decoupling.
This causes the relic neutrino temperature to be lower
by simple entropy considerations.  
A smaller relic neutrino energy density 
implies that larger neutrino degeneracies may be allowed. 
For example, we
have shown that interesting regions of the model parameters
for big-bang nucleosynthesis (BBN)
are allowed such that substantial lepton asymmetry and
baryon density  (even $\Omega_b h^2 \approx 1$ 
 where $h$ is the
present value of the Hubble constant in units of 100 km sec$^{-1}$ Mpc$^{-1}$) 
are possible while still satisfying the adopted  abundance 
constraints from primordial nucleosynthesis.  

A stronger constraint, however,
 on lepton asymmetry comes from the 
power spectrum of fluctuations in the cosmic microwave
background (CMB) which we now address in the present paper.    
We apply a likelihood analysis
of neutrino-degenerate models to the combined latest 
 BOOMERANG \cite{netterfield}, DASI \cite{pryke} and MAXIMA-1 \cite{hanany}
 results.    We note, however,
that a recent analysis \cite{dolgov01}
of the implications of neutrino oscillations
derived from a combination of the atmospheric, and solar neutrino constraints
implies much tighter limits on degeneracy for all neutrino flavors.  
If neutrino oscillation
parameters are in the range of the large mixing angle solution then an upper
limit of $\vert \xi_{\nu_i} \vert ^<_\sim 0.07$ applies to all neutrino flavors.
The limits derived here do not assume 
any particular  model for neutrino mixing, and should be taken as
independent of, and complementary to, those constraints.

The implications of the CMB data for neutrino-degenerate cosmologies
have been noted in a number of recent papers 
\cite{hannestad,lesgourgues00,lesgourguesm,kneller,esposito,hansen01,bowen,dolgov02}.
The constraints on the effective number of relativistic particles
can also arise in other contexts, such as cosmic quintessence
\cite{bean,yahiro}.
The present work, however,  differs from those in several respects.  
For one, we consider the most recent 
combined data sets,
not just the first year BOOMERANG data as in \cite{hannestad,lesgourgues00}.
and generate a marginalized likelihood function for the
neutrino degeneracy and other cosmological parameters.  
Many of the existing studies have marginalized
over a more limited set of cosmological parameters. For example,
in \cite{kneller} only $\omegab h^2$ and and neutrino degeneracy
were marginalized to set limits while other cosmological parameters were
set to various fixed values.
In \cite{lesgourgues00}, for example,
 no likelihood analysis was made.  In \cite{hannestad}
a likelihood analysis was made but without window functions.

For the present work we use the updated 
Radpack package \cite{bond} described below
which includes all relevant window functions.  
Another difference between our analysis and other works 
is that our marginalization utilizes 
a global minimum search algorithm  \cite{global}
rather than a discrete grid 
of cosmological parameters.  Marginalizing parameters for
each fixed value of one parameter requires at least 1000 model calculations
 to get $10^{-4}$ accuracy for 
the $\chi^2$ minima even by using this algorithm.
Nevertheless,  in this way we are sure to identify the true
marginalized likelihood functions.
 
The most similar recent likelihood analysis to that described here
is in the work of \cite{hansen01}.
Our analysis differs from \cite{hansen01}
in several respects. 
In the present work 
we make use of our deduced new family of baryon densities and lepton
asymmetries allowed by BBN to assign weak top-hat priors on the derived
likelihood functions. This differs from that in 
\cite{hansen01} in which separate 
Gaussian likelihood functions were evaluated
for the nucleosynthesis constraints and the CMB.  A total likelihood 
function was then defined by marginalizing over the product of these
two functions.  
 We prefer our method because the uncertainties in the BBN constraints 
are dominated by systematic errors.  Systematic errors are not equivalent
to random Gaussian errors. 
 We thus  prefer weak top-hat priors as a more realistic representation of the 
systematic errors in the BBN constraints. 

 One  other important difference is that we adopt a strong Gaussian
prior of $h = 0.72 \pm 0.08$ based upon the Hubble Key Project results
\cite{freedman}.
In \cite{hansen01} weak a top-hat prior of
$h = 0.65 \pm 0.20$ was  adopted.  
As noted above, another difference between the present work and 
all previous results
is that we consider carefully the change in background neutrino temperatures
as a function of degeneracy.  Although this is a small effect
for low degeneracies, it can slightly affect the upper limits.

\section{Neutrino-Degenerate BBN}
Neutrino degeneracy affects BBN in two ways.  The inclusion of
a small amount of electron $\nue$ degeneracy 
resets the equilibrium neutron to proton ratio at weak-reaction freezeout to
$n/p = \exp{\{-\Delta m/T - \xinue\}}$.  This can cause a reduction
in the primordial helium abundance.  Indeed, it has been argued
\cite{sato,kajino} that the  apparent conflict between 
the low helium abundance inferred from HII regions of metal poor galaxies
and  the low Lyman-$\alpha$ deuterium abundance may even require
\nue ~degeneracy for its resolution.
The present deuterium-absorption 
limits on $\Omega_b h^2 \approx  0.020 \pm 0.002~(2\sigma)$ requires a large 
primordial helium abundance of $Y_p~^>_\sim 0.25$  and substantial
destruction of primordial $^{7}$Li in stars.
These conditions tax even
the most generous adopted limits from observed light-element 
abundances \cite{steigman}.
Thus, a modification of BBN which allows for large values
of $\Omega_b h^2$ while still satisfying the constraints from 
light-element abundances is worth investigating.  
Such conditions are easily satisfied by neutrino-degenerate models.

The inclusion of either  \num ~or \nut~degeneracy 
on the other hand  only enhances the background energy density
and therefore the universal expansion rate.  
During the radiation dominated epoch,
relativistic neutrinos contribute a large fraction
of the mass energy.
Thus, even  a small modification of the neutrino energy density
can significantly affect the expansion.

The energy density  $\rho_\nu$ due to
degenerate neutrinos (or any other fermions)
are described by the usual 
Fermi-Dirac distribution functions 
$f_\nu =[\exp \left(E/T_\nu - \xi_\nu\right)+1]^{-1}$,
where the neutrino degeneracy parameter is defined  by
$\xi_{\nu} \equiv  \mu_{\nu}/T_{\nu}$, and  $\mu_{\nu}$
is the neutrino chemical potential.  Thus, we have
\begin{equation}
\rho_\nu + \rho_{\bar{\nu}} = 
{1 \over 2\pi^2}\int_0^\infty \!\! \! dp ~p^2 E_\nu
 (f_\nu(p)+ f_{\bar{\nu}}(p)).
\label{eq:rhonu}
\end{equation}
where, $p$ denotes the magnitude of the 3-momentum, 
and $E_\nu = \sqrt{p^2 + m_\nu^2}$, with $m_\nu$ the neutrino mass.
Here and throughout the paper we use natural units( $\hbar=c=k_B=1$).

For the present discussion it is sufficient to
only consider massless neutrinos.
(Possible limits on neutrino-degenerate models
with massive neutrinos are considered in \cite{lesgourguesm}).
The energy density in massless neutrinos becomes
\begin{equation}
\rho_{\nu}  + \rho_{\bar{\nu}} = {7 \over 8} {\pi^2 \over 15} \sum_i T_{\nu_i}^4
 \biggl[1 + {15 \over 7} \biggl(
{\xi_{\nu_i} \over \pi}\biggr)^4 + {30 \over 7} \biggl(
{\xi_{\nu_i} \over \pi}\biggr)^2\biggr]~~,
\end{equation}
from which it is clear that degeneracy in any neutrino species
tends to increase the energy density.
The  associated increased expansion rate tends to increase the
neutrino decoupling temperature. This causes
an increase in the primordial helium and other light-element abundances.

Since $\xinum $ and $\xinut$ primarily affect the expansion rate,
they are roughly interchangeable as far as their effects on 
nucleosynthesis or the CMB are concerned.  
Furthermore, it now seems likely \cite{numix}
that the mixing parameters
for $\nu_\mu$ and $\nu_\tau$ involve a large mixing angle and small
$\delta m^2$.  In this case it is plausible that the muon and tau neutrinos were
interconverted in the early universe and would therefore obtain
nearly an identical degeneracy parameter if an asymmetry exists.
Thus, we adopt a conservative model in which the $\mu$ and $\tau$ neutrinos
are equally degenerate,  
$\vert \xinum\vert =\vert \xinut\vert \equiv \xinumt$. 

\begin{figure}
\mbox{\psfig{figure=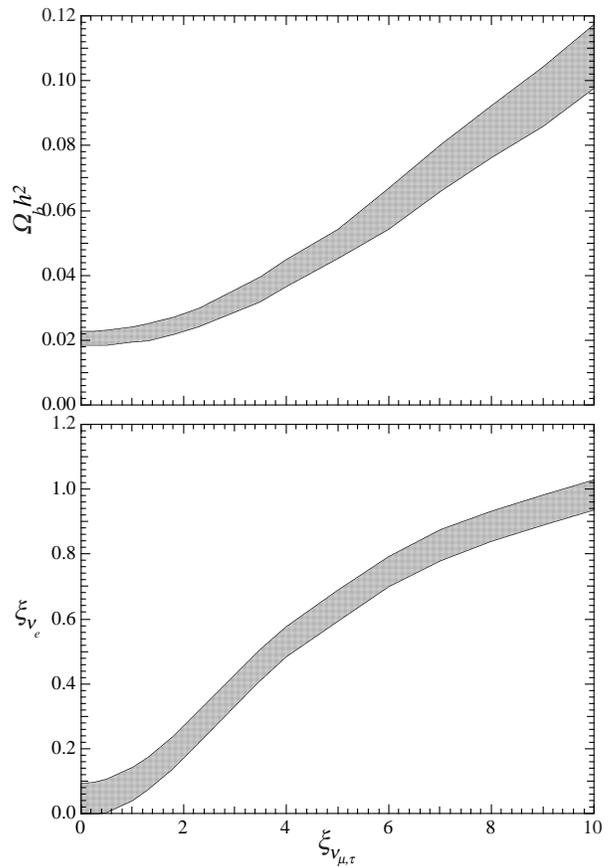,width=3.3in}}
\caption{Allowed values of $\Omega_b h^2$ and $\xinue$ for which the
constraints from light-element abundances are satisfied as a function
of $\xinumt$.
 }
\protect\label{fig:1}
\end{figure}

As shown in \cite{orito00}, for 
each value of $\xinumt$
there is a unique range of $\xinue$ and $\omegab h^2$ 
 which satisfies the combined
deuterium and primordial helium constraints.
The allowed family of neutrino-degenerate models employed in this
work is summarized in
Figure \ref{fig:1}. 

This figure differs slightly from the family of 
allowed solutions
given in \cite{orito00} in that we have adopted the newer
D/H constraint from \cite{omera} (i.e.~D/H $=3.0 (\pm 0.4) \times 10^{-5}$)
and slightly different limits on the primordial helium
abundance ($0.228 \le Y_p \le 0.248$).
In the limit of the standard nondegenerate big bang
($\xinumt = \xinue = 0$) our limits on $\omegab h^2$ reduce
to those of \cite{nollett00,tytler00}, i.e.~$\omegab h^2 = 0.021 \pm 0.002$.
The allowed shaded regions in Figure \ref{fig:1} will be adopted
as weak top-hat priors in the CMB likelihood analysis
described below. These regions include both the uncertainties from the
abundance constraints described above and the uncertainties in
the BBN model predictions \cite{cyburt}.

\section{CMB Power Spectrum}
Having defined the family of allowed priors from BBN
 we can now do a likelihood search for 
optimum  cosmological parameters which fit the CMB data.
Several recent works 
\cite{hu,lineweaver,kinney,lesgourgues,hannestad} have explained how
neutrino degeneracy can dramatically alter the power spectrum of the CMB.
For massless neutrinos it can be shown \cite{lesgourgues} that
the only effect of neutrino degeneracy is to increase the background
pressure and energy density of relativistic particles.
The essence of this constraint is that
degenerate neutrinos increase the energy density in
radiation at the time of photon decoupling and delay
the time of matter-radiation energy-density equality.
This mainly causes an increase in  the amplitude of the first
acoustic peak in the CMB power spectrum at $l \approx 200$.
For example, based upon a $\chi^2$ analysis 
\cite{lineweaver} of 19 experimental
points and window functions, it was concluded in \cite{lesgourgues}
that $\xi_\nu \le 6$ for a single degenerate neutrino species
with an $\Omega_\Lambda = 0$ cosmology.
   
However, in the  existing CMB constraint calculations 
\cite{hu,lineweaver,kinney,lesgourgues,hannestad,lesgourgues00,lesgourguesm,kneller,esposito,hansen01,bowen,dolgov02}
 only small degeneracy parameters with the standard relic neutrino
temperatures were studied in the derived constraint.
Hence, the possible effect of a diminished relic neutrino temperature 
at high degeneracy needs to be considered.  To investigate this
we have done calculations of the CMB power spectrum, 
$\Delta T^2 = l(l+1)C_l/2 \pi$
based upon the CMBFAST code of Seljak \& Zaldarriago \cite{cmbfast}.
We have explicitly modified this code to account for the 
contribution of massless degenerate neutrinos with varying
relic neutrino temperatures
$T_{\nu_i}$ for each species \cite{orito00}.  

The experimental uncertainties are non-Gaussian, but can be well
represented by an offset log-normal distribution \cite{bond}.
As in \cite{pryke} we have evaluated the $\chi^2$
goodness of fit for a range of theoretical power spectra $C_l$ 
as follows: We define the goodness of fit by
\begin{equation}
\chi^2 = \sum_{i,j} \bigl( Z_i^{t} - \bar Z_i^d\bigr) M_{i j}^Z \bigl(Z_j^{t}
 - Z_j^d\bigr) + \chi^2_{cal}~~,
\end{equation}
where separate summation over the different data sets is implied.
For each set of binned power data  $Z_i^{d}$ we utilize the published
off-set log-normal data from the three data sets.
\begin{equation}
Z_i^{d} \equiv  \ln{(D_i  + x_i)}
\end{equation}
Where $D_i$ is the measured band power.
The corresponding binned theoretical power spectra are
\begin{equation}
Z_i^{t} \equiv  \ln{\biggl[\sum_l \epsilon_i (W_{il}/l)
({\mathcal C}_l  + x_i )\biggr]}~~,
\end{equation}
where the $\epsilon_i$ are the published calibration uncertainties
 taken to be
8\%, for MAXIMA-1 and DASI, and 20\% for BOOMERANG.  
Window functions  $W_{i l}$ for the three data sets are available on the
world wide web.
The error matrix is simply 
\begin{equation}
M_{i j}^Z =  M_{i j}(D_i + x_i)(D_j + x_j)
\end{equation}
where $M_{i j}$ is the weight matrix for the band powers $D_i$.
The effect of the calibration uncertainty on the goodness of fit
is obtained from
\begin{equation}
\chi_{cal}^2 = \sum_i {(\epsilon_i -1)^2 \over \sigma_i^2}~~,
\end{equation}
where $\sigma_i$  is the experimental uncertainty.
The total  $\chi^2$ evaluated in this way can be converted \cite{pryke,lange}
 into a likelihood function for each parameter $x$ marginalized
over the remaining parameter set $\vec y$
\begin{equation}
{\mathcal{L}}(x) = \int P_{prior}(x,\vec y) exp{(-\chi^2/2)} d\vec y~~.
\end{equation}

In  neutrino-degenerate models which satisfy the constraints from
primordial nucleosynthesis \cite{orito00}, increasing the neutrino-degeneracy must
be accompanied by a commensurate increase in baryon density.
Fits to the CMB power spectrum for large degeneracy \cite{orito00},
therefore show a suppression of the  the second acoustic peak 
due to baryon drag \cite{hu}.

Indeed, such suppression of the second acoustic peak 
seemed to be present in the  first reported power spectra based upon the 
balloon-based CMB sky maps from the BOOMERANG \cite{boomerang}
and MAXIMA-1 \cite{hanany} collaborations.
This remains true for the likelihood analysis based upon a MAXIMA-1 data
\cite{balbi} which indicates $\Omega_b h^2 = 0.030^{+ 0.018}_{-0.010} ~(2\sigma)$.
 However, in the most recent data sets
from BOOMERANG \cite{netterfield} and DASI \cite{pryke}
the second peak has become much better defined.
Both the BOOMERANG and DASI  data  sets now  imply
$\Omega_b h^2 = 0.022^{+ 0.004}_{-0.003} ~(1 \sigma)$ ($\eta_{10} =
6.00^{+ 1.10}_{-0.81}$).   
This value is close to the value 
implied by the cosmic deuterium abundance
in high-redshift Lyman-$\alpha$ clouds observed
along the line of sight to background quasars
\cite{nollett00,tytler00}
$\Omega_b h^2 = 0.020 \pm 0.001~(1 \sigma)$ ($\eta_{10} =
5.46 \pm 0.27$).  Hence, the newer data imply at most a marginal
requirement for a larger baryon density
or neutrino degeneracy.  Indeed, these new data tighten
constraints on the possibility of degenerate cosmological neutrinos.
In the present paper we explore the new limits on possible
neutrino degeneracy implied by the combined data sets and our
BBN constraints.

\section{Results}
We limit our
consideration to flat $\Omega_{tot} = \Omega_M + \Omega_\Lambda = 1$ 
cosmological models with ionization parameter $\tau = 0$. 
This is sufficient for our
purposes since the likelihood functions so deduced are not 
expected to be much different if $\Omega_{tot}$ or $\tau$ are
varied (cf. \cite{netterfield}).  This is because $\tau$ and
the spectrum tilt $n$ are nearly degenerate parameters, i.e.~changing
one is equivalent to changing the other.  Moreover, $\Omega_{tot}$ is
 generally tightly constrained to be near unity anyway.

There are then nine parameters over which
we marginalize.  These  are $(\Omega_b h^2, \xinumt, \xinue
\Omega_\Lambda, h, n, \epsilon_i)$.
We utilize a strong Gaussian prior 
for $h = 0.72 \pm 0.08$ and for the calibration uncertainties
$\epsilon_i$ as listed above.
Also, as noted above, we adopt weak top-hat priors 
when marginalizing over $\Omega_b h^2$ and  $\xinue$
designated by the
shaded regions of Figure \ref{fig:1} for each value of $\xinumt$. 
In \cite{bowen} it has been argued that without some priors
on $\Omega_M$ (through flatness, $h$, etc.) it is difficult to
place bounds on the amount of relativistic matter.
Hence, the model constraints adopted here are probably required to
break the parameter degeneracy between relativistic and nonrelativistic
matter.  Ultimately, however, high resolution sky maps such as
the Planck mission  will be able to determine separately the amounts of relativistic
and nonrelativistic matter.

\begin{figure}
\mbox{\psfig{figure=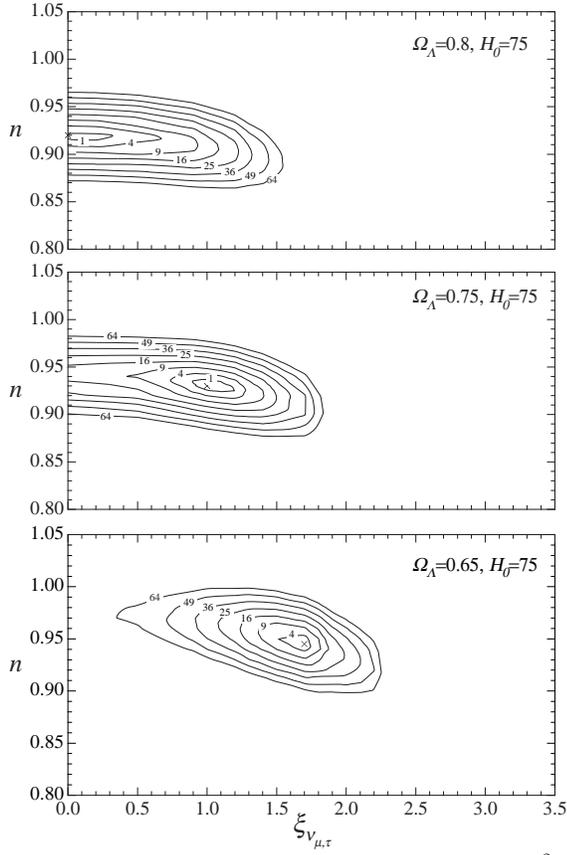,width=3.0in}}
\caption{Contours of constant goodness of fit $\Delta \chi^2$ in the 
$\xinumt$ vs. $n$ plane for three different $\Omega_\Lambda$ 
and $h = 0.75$ values as indicated.  Note the well developed minimum for
$\xinumt \approx 1-2$ and $\Omega_\Lambda \le 0.75$.
}
\protect 
\label{fig:2}
\end{figure}

Figure \ref{fig:2} illustrates one of the main results of this study.
Shown are contours of constant $\Delta \chi^2$  in the
$\xinumt$ vs.~$n$  plane for three 
values of the cosmological constant 
(i.e.~$\Omega_\Lambda = 0.65,~0.75,$ and $0.8$)
and for fixed $h = 0.75$ as noted.  
For $\Omega_\Lambda \le 0.75$, a  minimum
in $\chi^2$ develops for values of $\xinumt \approx 1-1.5$. 
Indeed, for a simple 2 parameter search with
 fixed values, $\Omega_\Lambda = 0.75$ and $h =0.75$,
 neutrino degeneracy is  preferred at the level of more than
$3 \sigma$ over a nondegenerate model.  
For smaller values of $\Omega_\Lambda$,
 this minimum for neutrino-degenerate
models becomes even more pronounced.  

A second minimum  also develops for higher degeneracy ($\xinumt \approx 11.4$)
as noted in \cite{orito00}.
This is due to the large change in particle degrees
of freedom for neutrinos which decouple just above the QCD
transition.  However, the goodness of fit 
is so poor ($\Delta \chi^2 \ge  500$) that it would not be
apparent in the contours drawn on Figure \ref{fig:2}. 
Hence, this large-degeneracy solution 
is definitely ruled out by the current CMB power spectrum. 

\begin{figure}
\mbox{\psfig{figure=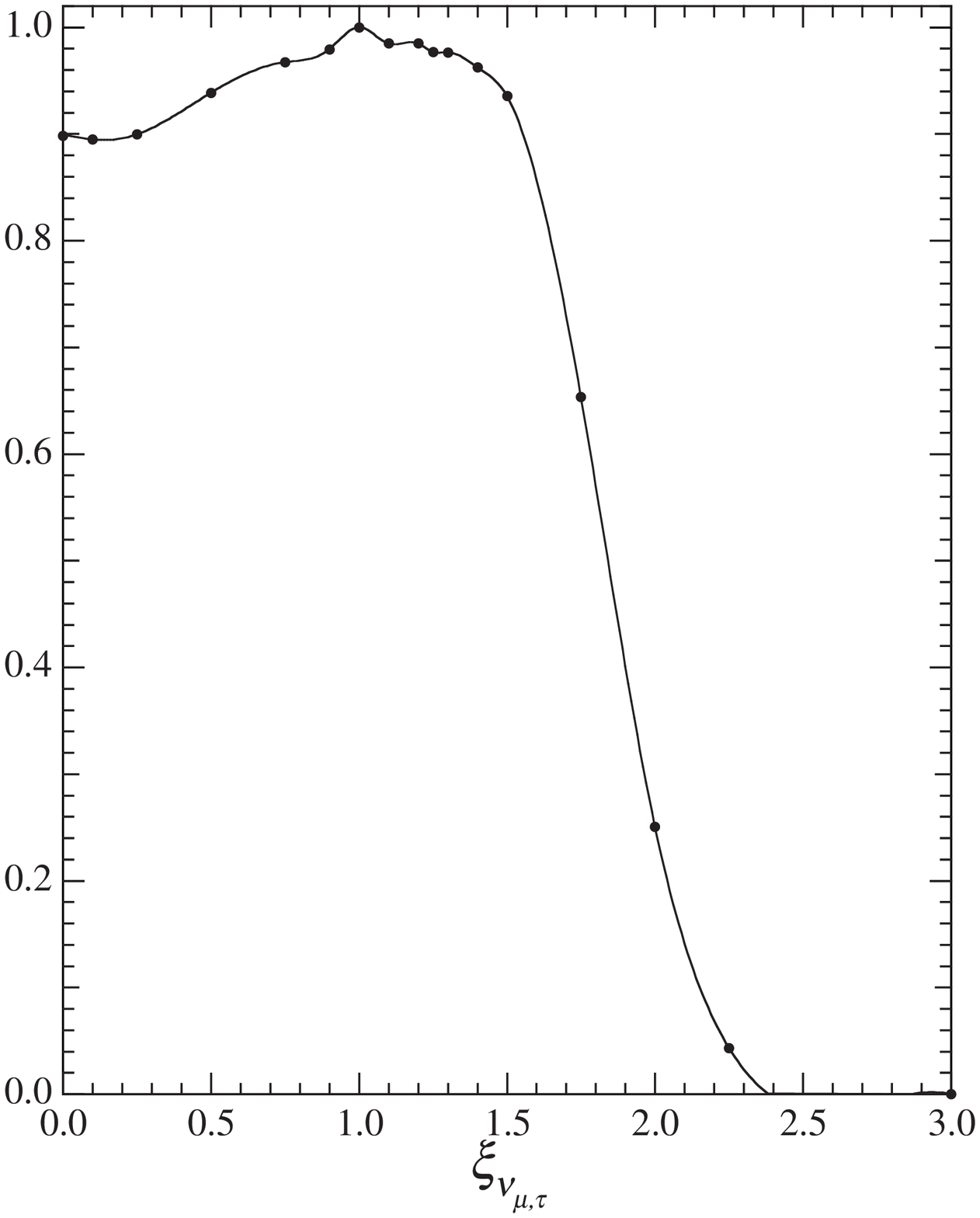,width=2.2in}}
\mbox{\psfig{figure=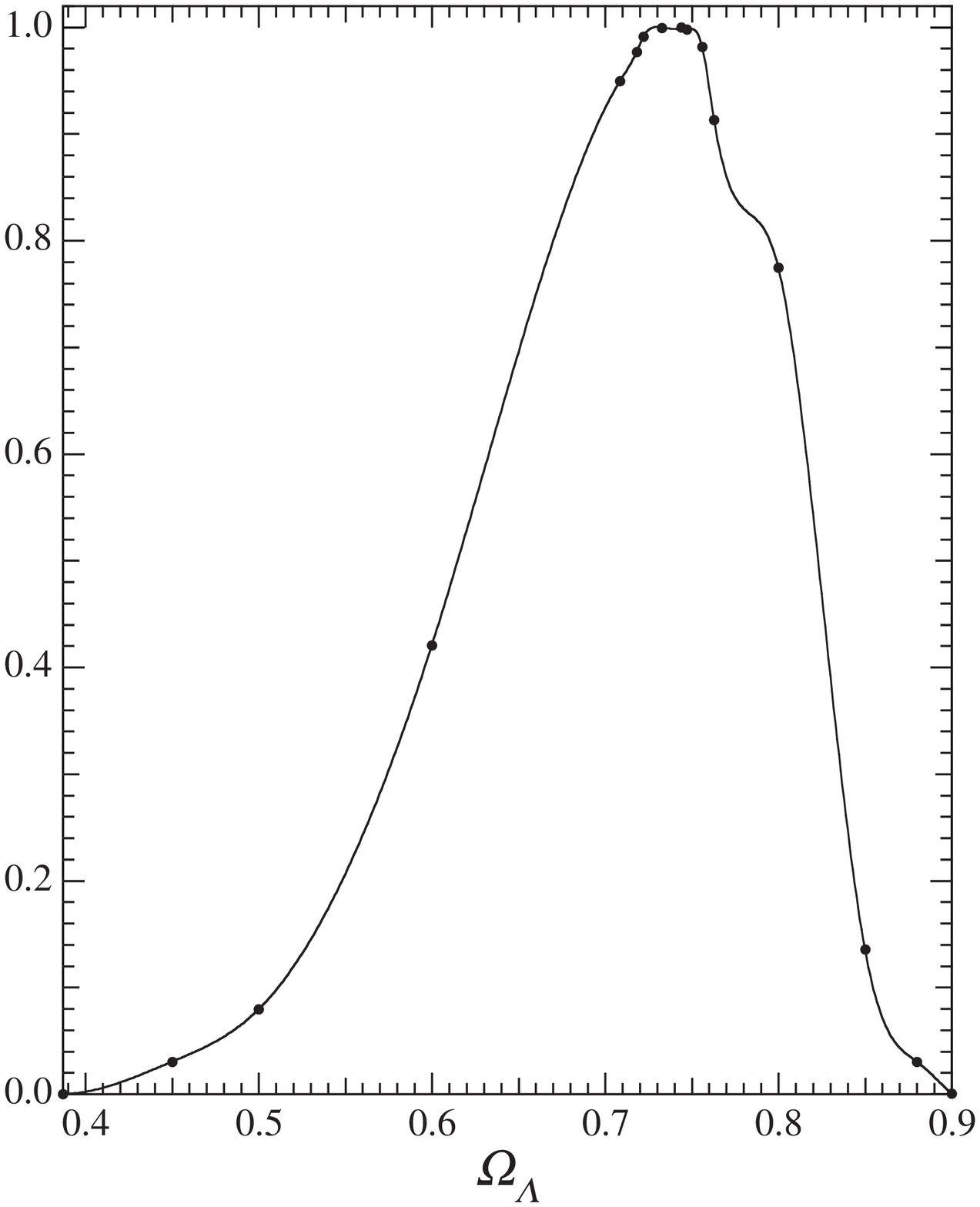,width=2.2in}}
\mbox{\psfig{figure=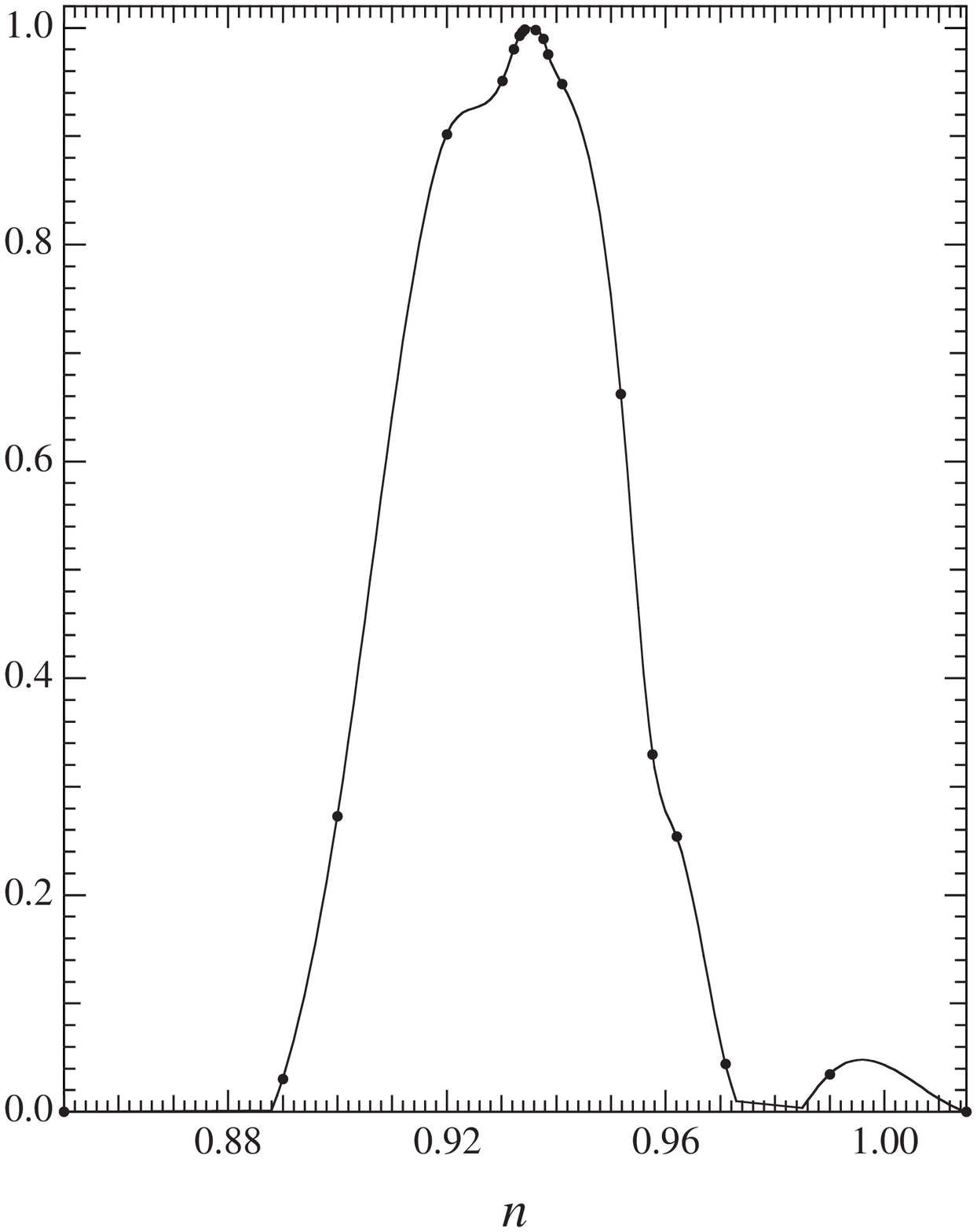,width=2.2in}}
\caption{Marginalized likelihood distribution functions for $\xinumt$,
$\Omega_\Lambda$, and the spectrum tilt $n$ as labeled.
}
\protect 
\label{fig:3}
\end{figure}

Figure \ref{fig:3} shows the marginalized likelihood distributions for three
of the cosmological parameters ($\xinumt,~\Omega_\Lambda$,~$n$) considered here.
For the present study
the likelihood functions  for $\xinue$, $\Omega_b h^2$ and $\Omega_M$
are related to these
since $\xinue$ and $\Omega_b h^2$ are functions of $\xinumt$ and 
$\Omega_M = 1 - \Omega_\Lambda$. 
From Figure \ref{fig:3} we deduce optimum values of 
$\Omega_\Lambda = 0.74^{+0.08}_{-0.11}$ 
and $n = 0.93 \pm 0.02$.  
A slight preference for finite neutrino degeneracy is 
evident $\xinumt = 1.0^{+0.8 (1\sigma)}_{-1.0 (0.5\sigma)}$.  
This preference, however,  is not particularly significant.
For now, the data mainly imply $(2 \sigma)$ upper limits
on neutrino degeneracy of $\xinumt \le 2.1$.  This value implies
upper limits of $\xinue \le 0.30$ and 
$\omegab h^2 \le 0.030$ from Figure \ref{fig:1}.
For a single large-degeneracy neutrino species, these limits become
$\vert \xinum \vert $ or $ \vert \xinut \vert  
 = 1.4^{+1.1 (1\sigma)}_{-1.4 (0.5\sigma)}$ with
a $2\sigma$ upper limit of $\xinumt \le 2.5$.

\begin{figure}
\mbox{\psfig{figure=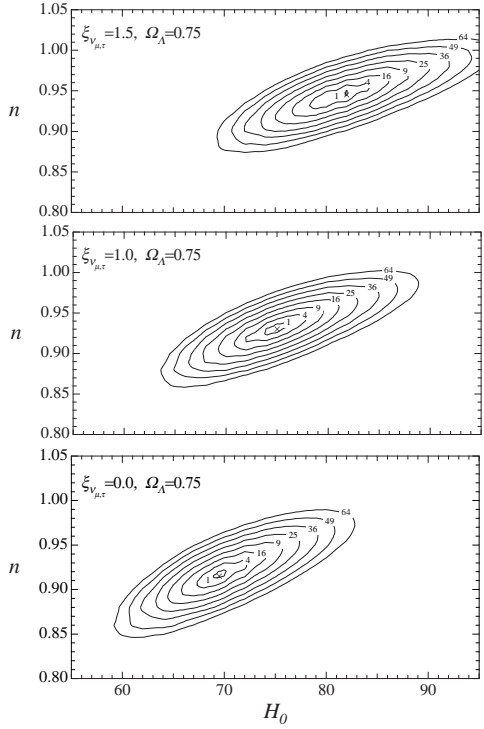,width=2.5in}}
\caption{ Contours of constant goodness of fit  $\Delta \chi^2$ in the 
$H_0$ vs.~$n$ plane for  $\Omega_\Lambda = 0.75$ and 
a neutrino degeneracy parameters $\xinumt = 0.,$ 1.0, and 1.5 as labeled.
}
\protect \label{fig:4}
\end{figure}

Our results are slightly more stringent than the results from
\cite{hansen01} who found an equivalent single species upper limit
based upon the CMB data alone 
of $\xinum$ or $\xinut < 2.9$.  
This is at first surprising given that we have adopted 
weak top-hat (instead of Gaussian) priors for the BBN constraint.
 We have traced  the main reason for the more stringent
upper limits derived here
to our adoption of a strong Gaussian prior on $h$.
A larger neutrino degeneracy
is possible if larger values of $h$ are permitted.
Figure \ref{fig:4} shows contours of constant $\Delta \chi^2$
in the $H_0$ vs.~$n$ plane  for $\Omega_\Lambda = 0.75$
models with $\xinumt = 0$, $1.0$  and $1.5$ as labeled.  
This illustrates the
sensitivity of the degenerate solution to the assumed prior
for $h$.  
If  weaker priors on $h$ are adopted, or
if new larger values of $h$ in the upper range of the 
present Key-Project uncertainty are ever determined,
the neutrino-degenerate models could become strongly preferred over the
non-degenerate models. 
The $\xinumt = 0$ non-degenerate solution is  only the preferred minimum,
for all values of $\Omega_\Lambda$,
when $h \le 0.70$.  This is  consistent with the results of \cite{balbi,lange}.

\begin{figure}
\mbox{\psfig{figure=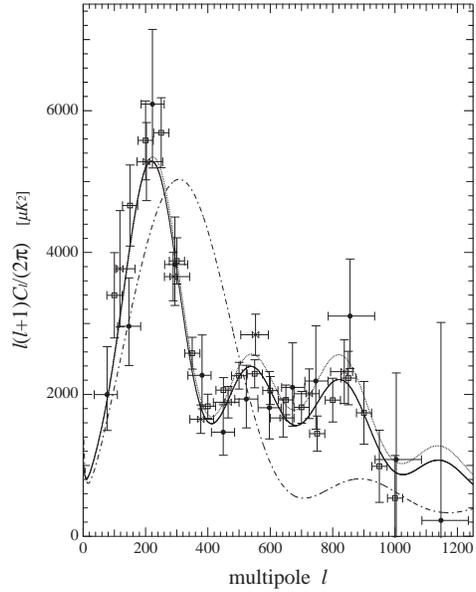,width=2.5in }}
\caption{ Fits to the power spectrum of fluctuations in the CMB.
The solid line shows the best neutrino-degenerate fit
$(\xinumt = 1.0)$.
The dotted line shows a best non-degenerate  ($\xinumt = \xinue = 0$)
 model.
For illustration, the dot-dashed line also shows
the large-degeneracy minimum $(\xinumt = 11.4)$.
}
\protect 
\label{fig:5}
\end{figure}

Figure \ref{fig:5} shows some 
optimum model power spectra  compared with the combined data sets. 
The  solid line shows our
optimum degenerate  model for which
($\omegab h^2$, $\xinumt$, $\xinue$, $\Omega_\Lambda$, $h$, $n$) $=$ 
($0.021$, $1.0$, $0.09$, $0.74$, $0.74$, $0.93$).
 For this parameter set we obtain a total 
$\chi^2 = 29.8$ for 29 degrees of freedom implying a nearly perfect fit.
For comparison, the
dotted line shows the best non-degenerate  ($\xinumt = \xinue = 0$)
 model [($\Omega_b h^2$, $\Omega_\Lambda$, $h$, $n$)  $=$  ($0.021$, $0.62$, $0.62$,$1.0$) (dotted line)]
from \cite{netterfield}.
For illustration we also show 
the large-degeneracy minimum~[($\Omega_b h^2$, $\xinumt$, $\xinue$,
$\Omega_\Lambda$, $h$, $n$) $=$ ($0.052$, $11.4$, $0.74$,
$0.45$, $0.80$, $0.72$) (dot-dashed line)].

\section{Conclusions}
In neutrino-degenerate models the larger 
baryon density associated with the observed low deuterium abundance
can be more easily accommodated than in non-degenerate models. 
Moreover, neutrino-degenerate models
provide a slightly improved 
goodness of fit for the latest CMB power spectra from
BOOMERANG, DASI, and MAXIMA-1.

Using cosmological models consistent with
the constraints from light-element abundances
as a function of the neutrino degeneracy parameter $\xinumt$,
we have shown that 
a slight maximum in the likelihood function
forms for neutrino-degenerate models with
$\xinumt \approx 1$.
However, the improvement over the nondegenerate models 
 is only at the level of about
0.5$\sigma$.
Although this minimum is not particularly statistically significant
for the present data set and assumed priors, 
it could become much more pronounced should larger
values of $h$ and/or smaller values of $\Omega_\Lambda$ ever be established
near their current $1\sigma$ limits.

The present data place $2\sigma$ 
limits for two identical large-degeneracy neutrino species is $\xinumt \le 2.1$,
which implies $\xinue \le 0.30$.
For only one species with large degeneracy, the limit becomes
$\vert \xinum \vert $ 
or  $\vert \xinut \vert \le 2.5$.
This is slightly
more restrictive than the limits deduced in other studies.

Finally, we remark that, since neutrino degeneracy is now 
limited to such small values, the present work has established that 
the effects of the changing neutrino decoupling
temperature with increased degeneracy has little effect.
Hence, previous studies which neglected this effect are justified.

\acknowledgments
One of the authors (GJM) wishes to acknowledge the hospitality
of the National Astronomical Observatory of Japan where some of this
work was done.
This work has been supported in part by the Grant-in-Aid for
Scientific Research (10640236, 10044103, 11127220, 12047233, 13002001,13640313) 
of the Ministry of Education, Science, Sports, and Culture of Japan.
Work at Univ.~Notre Dame supported 
by DoE Nuclear Theory Grant (DE-FG02-95-ER40934.
Work at Univ.~Oklahoma supported by NSF CAREER grant AST-0094335.

%

%
%
\vfill\eject
\vfill\eject

\end{document}